\documentclass{jaa}
\usepackage{natbib}
\usepackage{graphicx}
\usepackage{xcolor}
\usepackage{soul}

\begin{document}\sloppy

\title{Spectroscopy of 9 eruptive young variables using {\it  TANSPEC}}


\author{Arpan Ghosh\textsuperscript{1,3,*}, Saurabh Sharma\textsuperscript{1}, Joe P. Ninan\textsuperscript{2}, Devendra K. Ojha\textsuperscript{2}, A. S. Gour\textsuperscript{3},
Rakesh Pandey\textsuperscript{4}, Tirthendu Sinha\textsuperscript{5}, Aayushi Verma\textsuperscript{1}, Koshvendra Singh\textsuperscript{2}, Supriyo Ghosh\textsuperscript{6} and Harmeen Kaur\textsuperscript{7} }
\affilOne{\textsuperscript{1}Aryabhatta Research Institute of Observational Sciences (ARIES)  Manora Peak, Nainital 263 001, India.\\}
\affilTwo{\textsuperscript{2} Department of Astronomy and Astrophysics, Tata Institute of Fundamental Research (TIFR), Mumbai 400005, Maharashtra, India. \\}
\affilThree{\textsuperscript{3}School of Studies in Physics and Astrophysics, Pandit Ravishankar Shukla University, Raipur 492010, Chhattisgarh, India.\\}
\affilFour{\textsuperscript{4} Physical Research Laboratory, Navrangpura, Ahmedabad - 380 009, India.\\}
\affilFive{\textsuperscript{5} Satyendra Nath Bose National Centre for Basic Sciences (SNBNCBS), Salt Lake, Kolkata-700 106, India. \\ }
\affilSix{\textsuperscript{6} University of Hertfordshire, Hatfield, AL10 9AB, United Kingdom.\\}
\affilSeven{\textsuperscript{7} Center of Advanced Study, Department of Physics DSB Campus, Kumaun University, Nainital 263001, India.\\ }


\twocolumn[{

\maketitle

\corres{arpan@aries.res.in}


\begin{abstract}

In recent times, 3.6m Devasthal Optical Telescope (DOT) has installed an optical to near infra-red 
spectrograph, TANSPEC, which provides spectral coverage from 0.55-2.5 microns. Using TANSPEC, we
have obtained a single epoch spectrum of a set 9 FUors and EXors. We have analysed line profiles of
the sources and compared with the previously published spectra of these objects. Comparing the 
line profile shapes with the existing theoretical predictions, we have tried to interpret the 
physical processes that are responsible for the current disc evolution and the present accretion
dynamics. Our study has shown the importance for time evolved spectroscopic studies for better
understanding the evolution of the accretion mechanisms. This in turn can help in better 
categorisation of the young stars displaying episodic accretion behaviour.

\end{abstract}


}]


\doinum{12.3456/s78910-011-012-3}
\artcitid{\#\#\#\#}
\volnum{000}
\year{0000}
\pgrange{1--}
\setcounter{page}{1}
\lp{1}

\section{Introduction}\label{pt1}

\begin{figure*}
\centering
\includegraphics[width=0.45\textwidth]{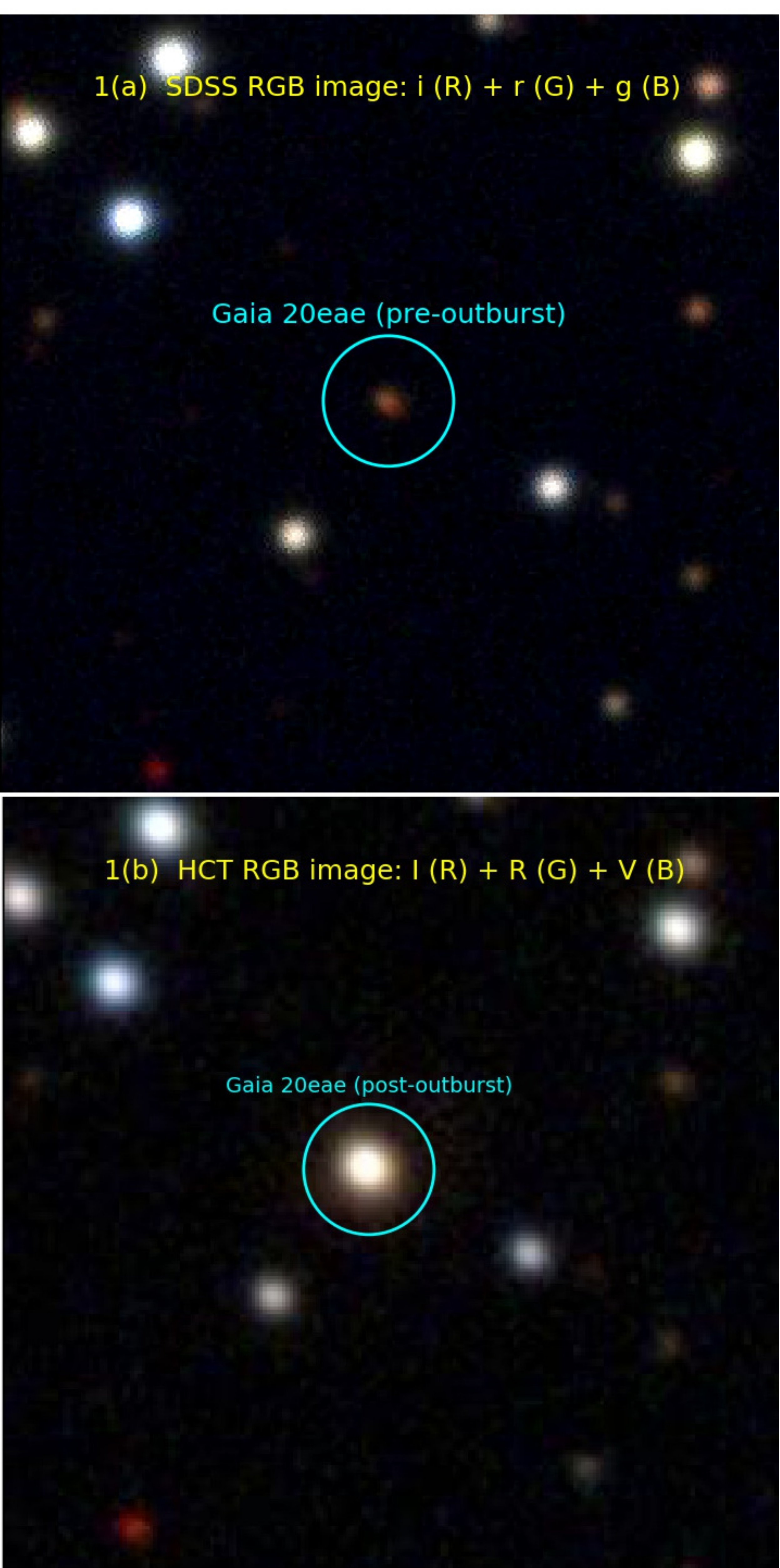}
\caption{\label{ref_img1} Optical RGB pre-outburst and 
post-outburst images of Gaia 20eae, a newly discovered 
episodically accreting young star. The pre-outburst RGB  
is generated from the g,r \& i filter images obtained from 
SDSS archive. Post-outburst RGB is generated from optical observations in V,R \& I 
filters taken from the 2m Himalayan {\it Chandra} Telescope.
Image source: \cite{2022ApJ...926...68G} }
\end{figure*}

The phenomenon of episodic accretion first came to light in 1936 when FU Ori located in the B35 dark clouds in the Orion star forming region underwent an outburst of more than 5 magnitudes 
from its historical V band magnitude of 16 \citep{1954ZA.....35...74W}. 
\citet{1966VA......8..109H} first proposed that this phenomenon of episodic accretion represents an important stage 
in the formation of low mass stars (mostly having masses $\leq$ 1 M$_{\odot}$ with few exceptions with masses between 2-3 M$_{\odot}$, \citealt{2016ARA&A..54..135H}). Observationally it has 
been found that the phenomena of
episodic accretion span the entire pre-main sequence (PMS) evolutionary stage starting from Class {\sc 0} to Class {\sc II} stages \citep{2015ApJ...800L...5S}. The time span of these events is
extremely short ($\sim$ months to $\sim$ decades) compared to the millions of years spent in the formation stages makes these events extremely rare with around 30 sources have been discovered
till date. These events though very small in timescales, are capable of delivering significant amount of matter onto the central source  \citep{2006ApJ...650..956V}. The brightness variations
exhibited in an episodic accretion event is easily noticeable in observations taken before and during the outburst. Figure \ref{ref_img1} shows one such example with the pre-outburst and post-outburst optical images of Gaia 20eae, a newly discovered member of the young stars displaying episodic accretion behaviour.
Sources displaying episodic
accretion are bi-modally classified into FUors and EXors. FUors named after FU Ori, undergo outburst of 4-5 magnitude in optical wavelengths that lasts for several decades to even $\sim$ 100
years and show absorption features in their spectra. EXors, named after Ex Lupi, on the other hand undergo smaller magnitude outbursts of 2-3 magnitude in optical lasting for few years and
display emission features in their spectra \citep{1996ARA&A..34..207H,1998apsf.book.....H}. 

Various models have been put forward to explain the enhancement of the accretion rate leading to episodic accretion. \citet{2009ApJ...701..620Z} proposed the throttle mechanism
to explain the outbursts. According to this model, the disc is replenished from the infalling envelope at a different rate than the rate at which the matter is transported to its inner regions 
resulting in the pile up of material which is subsequently released causing outburst. Another model, initially proposed by \citet{1994ApJ...427..987B} and later improved upon by 
\citet{2001MNRAS.324..705A}, postulates that the phenomena of episodic accretion events is self regulated. According to this theory, if the inward transport of matter within the disc is 
sufficiently high, it results in the ionisation of the gas (mainly hydrogen) causing rapid rise in opacity. This leads to heat being trapped within the disc that develops into a runaway situation during which the inner disc is emptied out. 
The third model proposes about the perturbation of the circumstellar disc by an
eccentric binary companion at periastron \citep{1992ApJ...401L..31B,2004ApJ...608L..65R}. Fourth scenario, envisages the idea of accretion of a large ``gas blob'' from the circumstellar disc to the central source \citep{2005ApJ...633L.137V}. The fifth and final
model is centred around the idea of disc-magnetosphere interactions in ``weak propeller'' regime to explain the weak ``EXor type'' outbursts \citep{2010MNRAS.406.1208D,2012MNRAS.420..416D}.  


Spectroscopic studies of sources exhibiting episodic accretion are 
crucial as the line profile shapes offer vital clues about the 
physical processes that are undergoing at their current stage of 
evolution. The sources classified as FUors and EXors broadly affirms 
their respective spectroscopic behaviour, but their photometric 
behaviour is quite diverse with different rise and decay timescales 
to and from their maximum brightness. Lately, several sources have 
been discovered which show intermediate behaviour to that of FUors 
and EXors \citep{2014prpl.conf..387A}.  In recent times,
{\it Gaia} photometric alert system has provided invaluable 
contribution towards identifying new eruptive young stars \citep{2013RSPTA.37120239H}. From the Gaia alerts, four new members 
of eruptive variables has been discovered: Gaia 17bpi \citep{2018ApJ...869..146H}, Gaia 18dvy \citep{2020ApJ...899..130S}, 
Gaia 19ajj \citep{2019AJ....158..240H} and Gaia 20eae \citep{2022ApJ...926...68G}. Among  them, Gaia 17bpi and Gaia 18dvy 
exhibited properties similar to FUors
whereas Gaia 19ajj resembled EXor.
Gaia 20eae displayed intermediate behaviour to that of FUors and EXors. 

In this paper, we discuss our findings based on a set of 9 eruptive 
young variables that were classified either as FUors or EXors \citep{2014prpl.conf..387A}. These sources are well studied with
their optical and near-infrared (NIR) spectra available in the literature. Recently, TIFR-ARIES Near Infrared Spectrometer \citep[TANSPEC,][]{2022PASP..134h5002S} was commissioned on the 3.6m 
telescope at Devasthal, Nainital, India \citep{2018BSRSL..87...29K}. 
This instrument is having a very unique capability to do  medium 
resolution (R$\sim$2750) spectroscopy from optical to NIR bands in a 
single shot. Thus, observing eruptive young variables through 
TANSPEC, we can compare their spectral properties both in optical and NIR 
bands with the previous published spectra. This is very 
crucial to  understand the physical processes responsible for the 
episodic accretion and their evolution. 
The paper is arranged as follows: Section \ref{pt2} describes about 
the observation and data reduction details. Section \ref{pt3} 
describes the results and we conclude this paper based on our results 
in Section \ref{pt4}.

\begin{table*}
\centering
\small
\caption{Log of Photometric and Spectroscopic Observations.}
\label{tab:obs_log}
\begin{tabular}{@{}rrrrrr@{}}
\hline 
Telescope/Instrument & Date  & JD & Source Name  & slit size & Exposure(sec)  \\
\hline 
3.6m DOT TANSPEC & 2020 Nov 12 & 2,459,166 & FU Ori      & 0.$^\prime$5 & 150 sec $\times$ 4 frames  \\
3.6m DOT TANSPEC & 2020 Oct 23 & 2,459,146 & XZ Tau      & 0.$^\prime$5 & 120 sec $\times$ 4 frames  \\          
3.6m DOT TANSPEC & 2020 Oct 24 & 2,459,147 & V2492 Cyg   & 0.$^\prime$5 & 150 sec $\times$ 8 frames   \\
3.6m DOT TANSPEC & 2020 Nov 12 & 2,459,166 & V2494 Cyg   & 0.$^\prime$5 & 150 sec $\times$ 6 frames    \\
3.6m DOT TANSPEC & 2020 Nov 10 & 2,459,164 & V1180 Cas   & 1.$^\prime$0 & 150 sec $\times$ 8 frames     \\
3.6m DOT TANSPEC & 2020 Nov 10 & 2,459,164 & V899 Mon    & 1.$^\prime$0 & 150 sec $\times$ 3 frames      \\
3.6m DOT TANSPEC & 2020 Oct 29 & 2,459,152 & V960 Mon    & 0.$^\prime$5 & 150 sec $\times$ 6 frames       \\
3.6m DOT TANSPEC & 2020 Oct 29 & 2,459,152 & V1118 Ori   & 0.$^\prime$5 & 180 sec $\times$ 12 frames       \\
3.6m DOT TANSPEC & 2020 Oct 29 & 2,459,152 & V733 Cep    & 0.$^\prime$5 & 180 sec $\times$ 12 frames       \\
\hline
\end{tabular}
\end{table*}

\section{Observation \& Data reduction}\label{pt2}

\begin{figure*}
\centering
\includegraphics[width=0.95\textwidth]{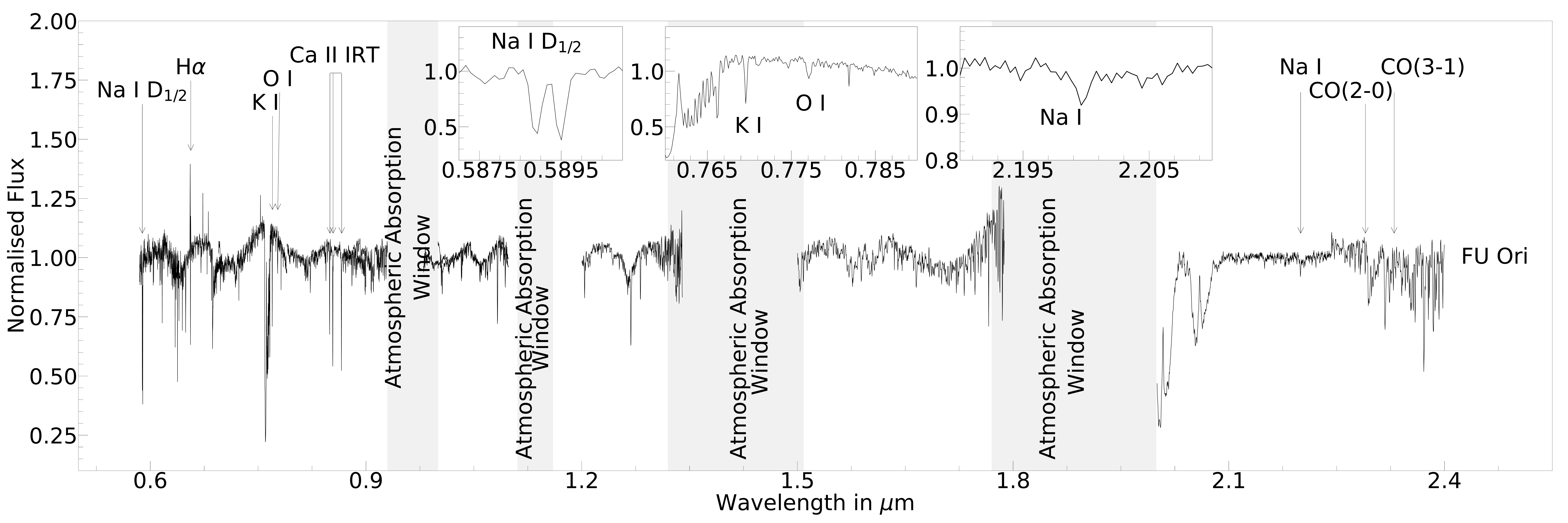}
\includegraphics[width=0.95\textwidth]{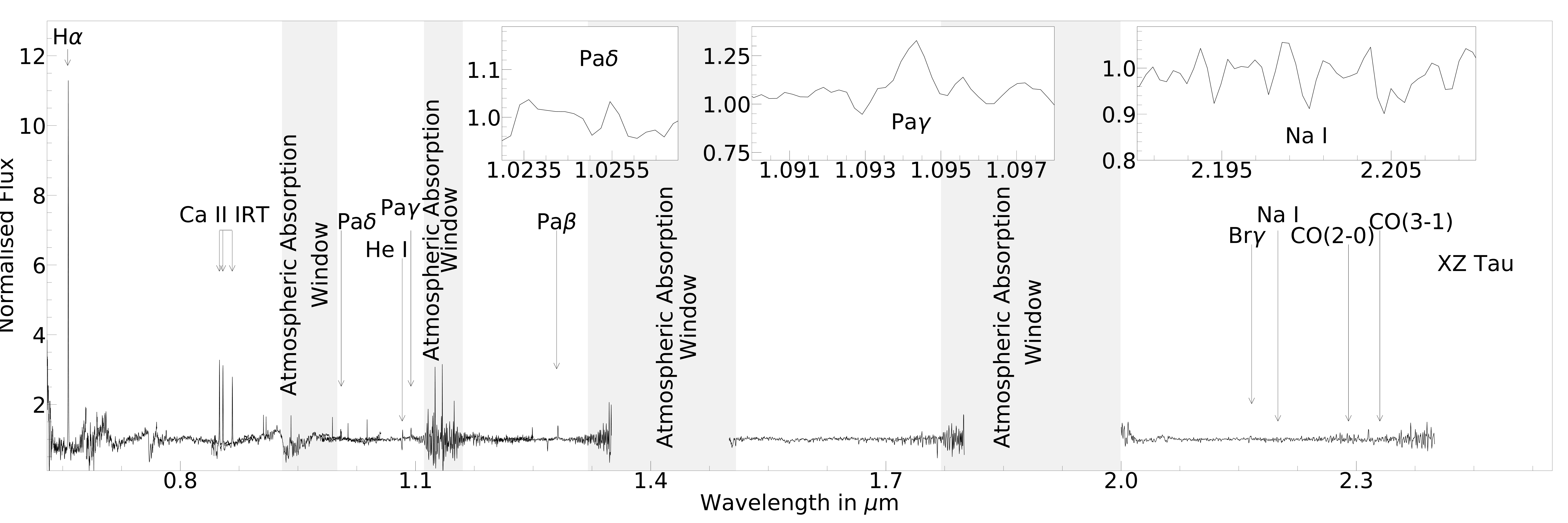}
\includegraphics[width=0.95\textwidth]{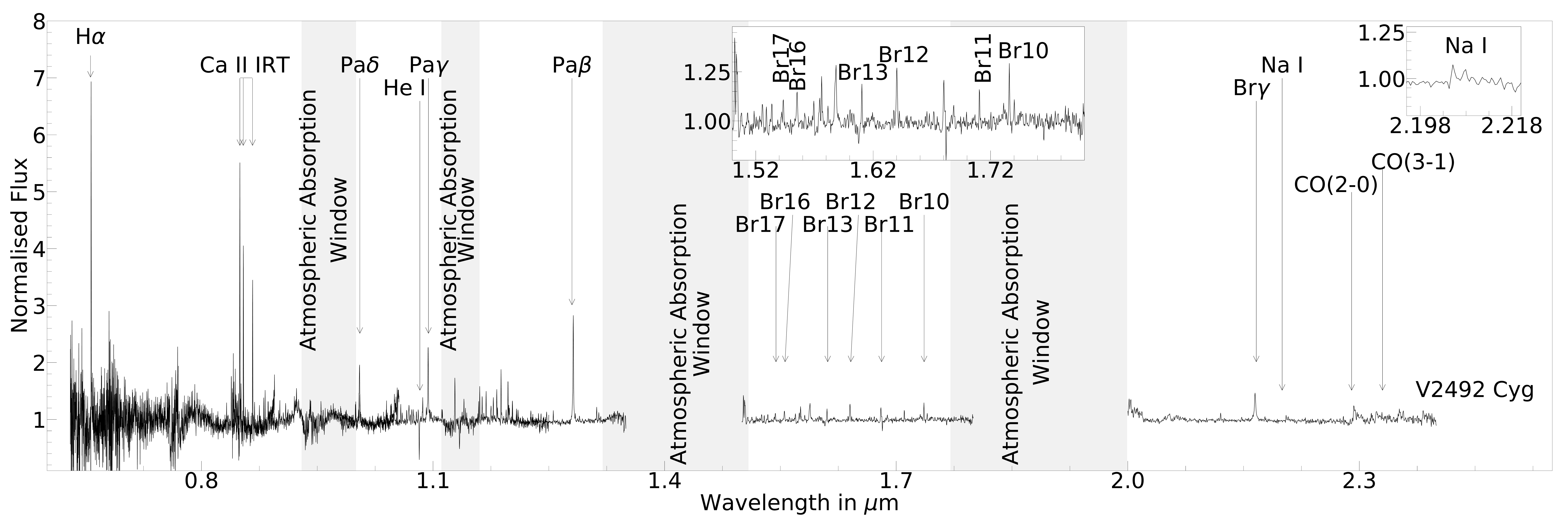}
\includegraphics[width=0.95\textwidth]{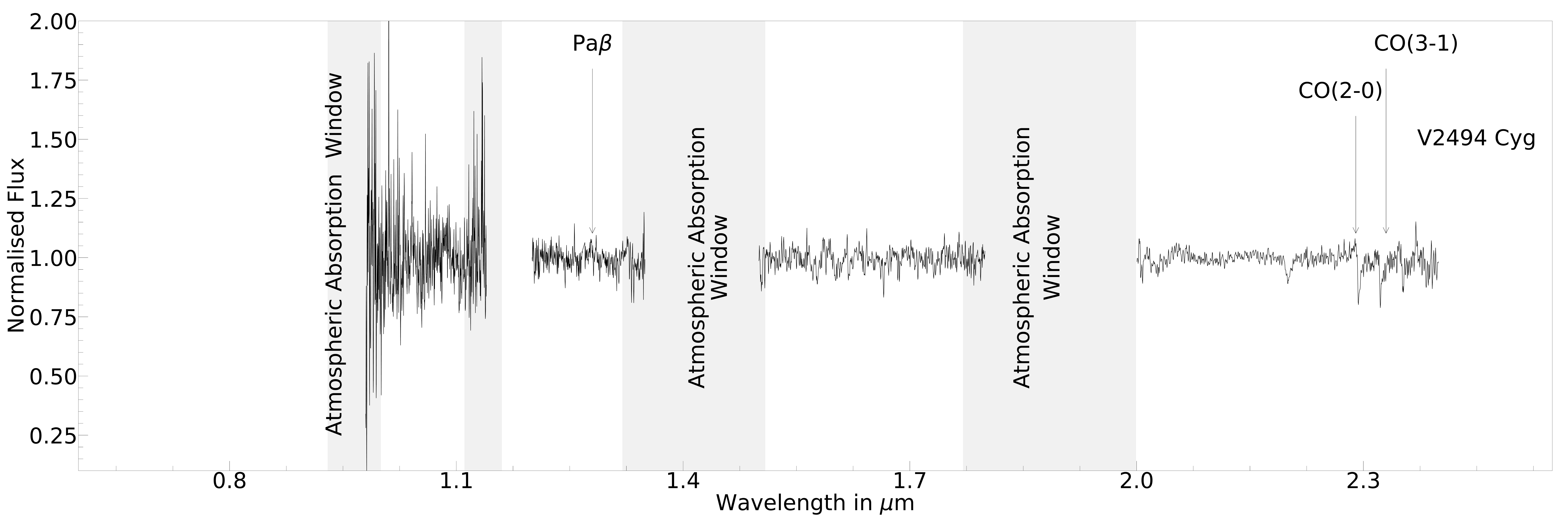}
\caption{\label{atlas_img1} Spectra of the young eruptive variables observed with {\it TANSPEC}. The 
lines which has been discussed in the present study have been marked. The Paschen and Brackett 
series lines are abbreviated as ‘Pa’ and ‘Br’ respectively. The shaded grey regions denote the 
atmospheric absorption window in our spectrum.}
\end{figure*}

We have carried out our observations using the 3.6m Devasthal Optical Telescope (DOT) located at 
Devasthal, Nainital (latitude =29$^\circ$21$^\prime$39$^{\prime\prime}$.4 N, 
longitude= 79$^\circ$41$^\prime$03$^{\prime\prime}$.6 E, altitude = 2450 m) with the newly 
commissioned {\it TANSPEC} instrument \citep{2022PASP..134h5002S}. {\it TANSPEC} has two spectroscopic modes : higher resolution 
cross-dispersed mode (XD) and a lower resolution prism mode, having a simultaneous wavelength coverage from optical to near-infrared bands, i.e., 0.55-2.5 $\mu$m. We have taken our observations with the
XD mode. The XD mode provides a median resolution (R) of $\sim$ 2750 across the orders 
in 0$^\prime{^\prime}$.5 slit and about 1500 in 1$^\prime{^\prime}$.0 slit. We have used both the 0$^\prime{^\prime}$.5 and the
1$^\prime{^\prime}$.0 slit for our observations. Table \ref{tab:obs_log} contains the list of objects
that we have observed along with the slit information and the exposure time. We have followed the 
standard observing procedures while obtaining spectra of each source. Each source was nodded 
along the slit at two positions and multiple exposures of the source at each position were taken to obtain the spectra. 
Our observations 
strategy were driven by the motto of obtaining high signal-to-noise-ratio (SNR) spectrum of our sources.
The maximum exposure time given to obtain one spectrum was limited to three minutes. This was done 
so as to cancel out the telluric emission lines from the consecutive frames taken at alternate nod
positions. We have also observed a nearby telluric standard star of A0V spectral type for telluric 
correction. We have also observed the argon and neon lamps for wavelength calibration and the
tungsten lamps for flat fielding. These calibration and arc lamps were observed for each source 
and target. The obtained spectrum is then processed, extracted and flat-fielded using the pyTANSPEC
pipeline\footnote{https://github.com/astrosupriyo/pyTANSPEC} developed for reducing the {\it TANSPEC} XD 
mode spectrum. The obtained output spectrum is then telluric corrected using the standard 
IRAF\footnote{IRAF is distributed by National Optical Astronomy
Observatories, USA which is operated by the Association of Universities for Research in Astronomy, Inc., under cooperative agreement with National Science Foundation for performing image processing.} module of {\it SPLOT}. After telluric correction, the resultant output was normalised using the 
{\it CONTINUUM} module of IRAF. Finally the different orders of continuum normalised spectra are
added up to generate the total spectrum of a single source. 
Spectra of each of the observed source is provided in Figure \ref{atlas_img1}. Various absorption and emission features are seen. There are gaps in our spectrum which is due to the atmospheric absorption windows due to the broad H$_2$0 and OH bands.

\setcounter{figure}{1}
\begin{figure*}
\centering
\includegraphics[width=0.95\textwidth]{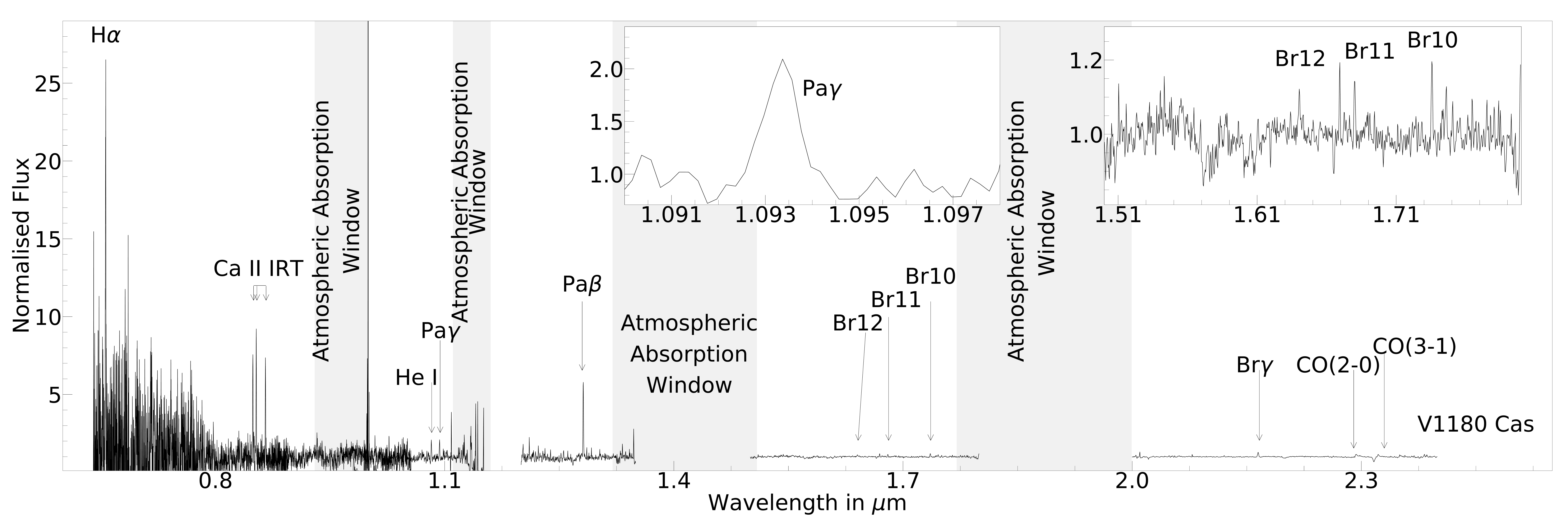}
\includegraphics[width=0.95\textwidth]{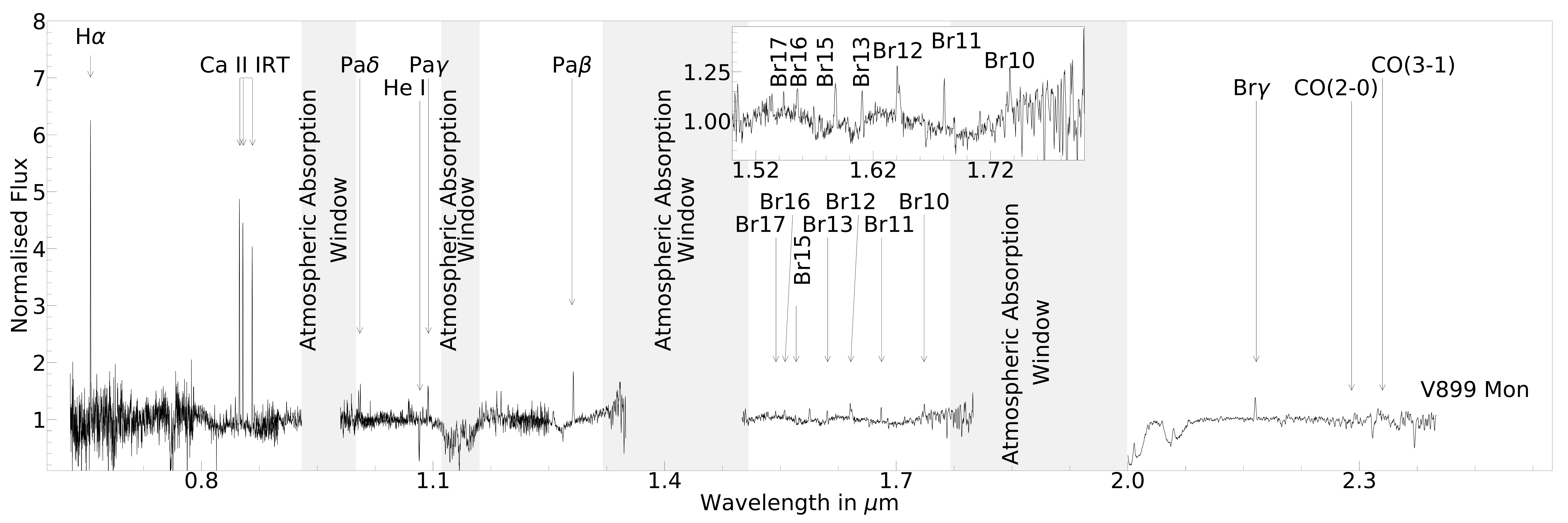}
\includegraphics[width=0.95\textwidth]{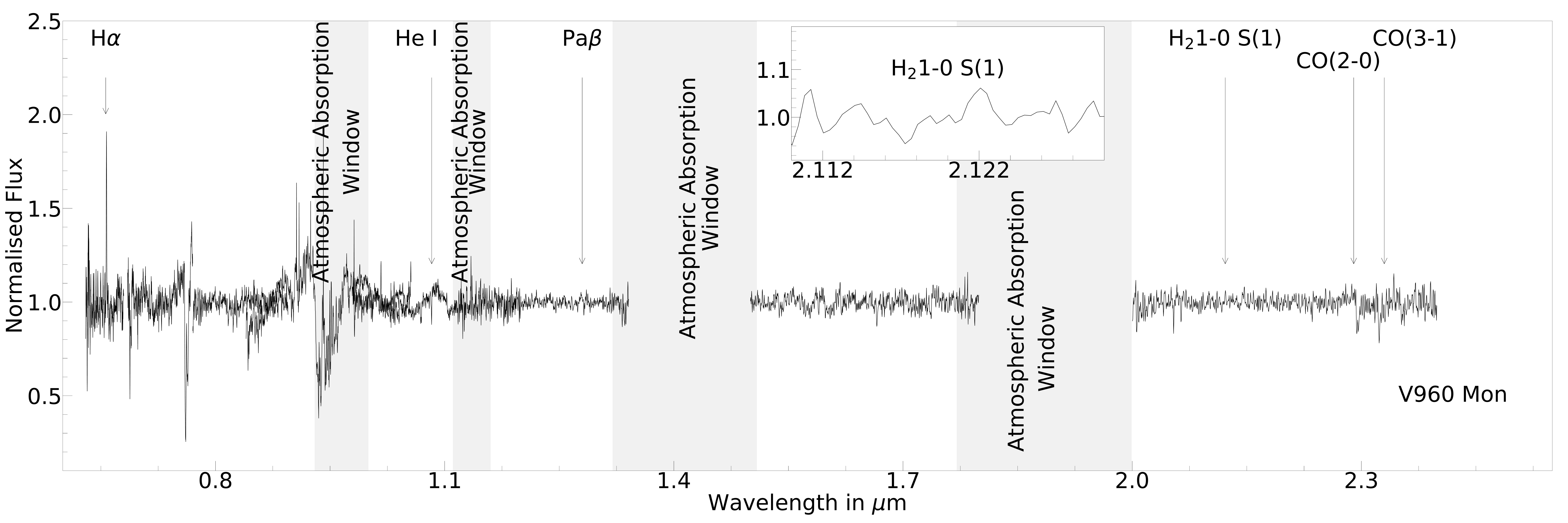}
\includegraphics[width=0.95\textwidth]{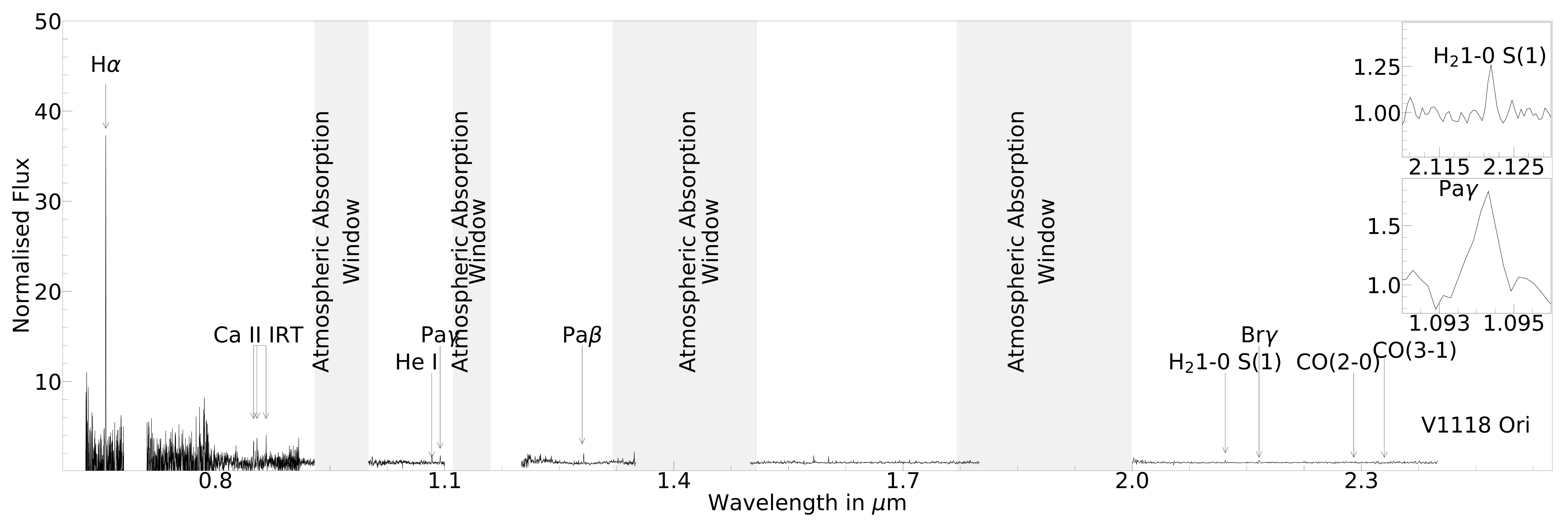}
\caption{\label{atlas_img2} Contd.}
\end{figure*}

\setcounter{figure}{1}
\begin{figure*}
\centering
\includegraphics[width=0.95\textwidth]{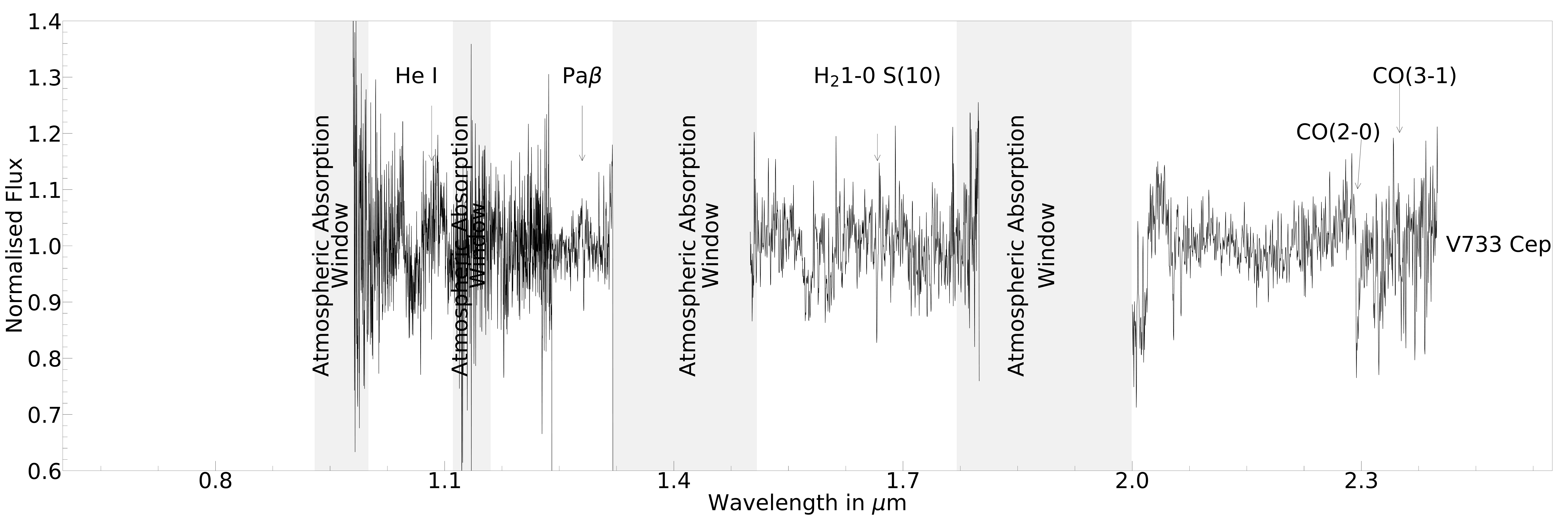}
\caption{\label{atlas_img3} Contd.}
\end{figure*}

\begin{figure*}
\centering
\includegraphics[width=0.95\textwidth]{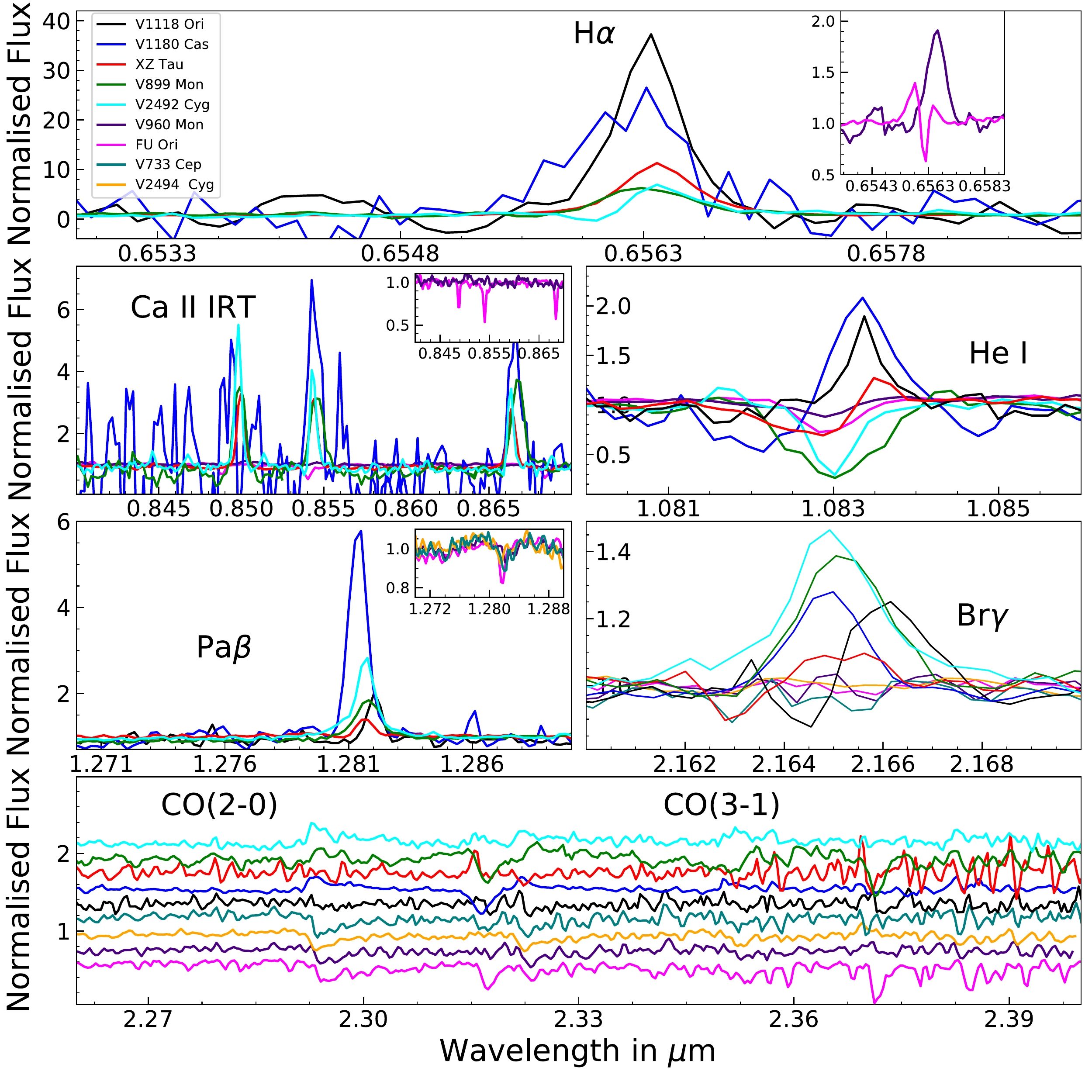}
\caption{\label{atlas_img4} Zoomed line profile of the important lines that are discussed in this paper.}
\end{figure*}

\section{Results \& Discussion} \label{pt3}

In the following sub-sections, we will be discussing the properties of the optical to near-infrared spectral lines found in the {\it TANSPEC} spectra (cf. Figures \ref{atlas_img1}) of our studied sample of FUors/EXors. Figure \ref{atlas_img4} shows the zoomed in version of the most common lines that are discussed in this paper.

\subsection{\bf FU Ori}

FU Ori is the first observed member of the class of young stellar objects (YSOs) displaying episodic accretion behaviour 
as it underwent a brightening of more than 5 magnitudes 
from its historical magnitude of $\sim$ 16 magnitude in V band in the year 1936 \citep{1954ZA.....35...74W}. It has since then 
faded by $\sim$ 1/2 magnitudes in the next 40 years \citep{1977ApJ...217..693H}. It is located in the
B35 dark nebula present in the Orion star forming region. FU Ori, being the first prototype of the 
category of episodic accreting YSOs and that of FUor subgroup, its spectrum serves as a reference for
all other potential new episodically accreting YSOs and FUors in general.

{\bf CO bandheads in K:}
Spectroscopically, FUors display an absorption feature while EXors exhibit an emission feature in
the (2-0) and (3-1) CO bandheads. 
FU Ori, being the first prototype of FUors display  deep absorption
bands in  CO bandheads. The origin of such deep absorption feature in CO bandheads along with the absence
of Br$\gamma$ emission in the spectrum can be attributed to the 
boundary layer accretion mechanism, which is different from magnetospheric accretion regime. In the
boundary layer accretion model, the accretion rate is so high that the ram pressure of the 
inflowing matter in the disc breaks through the magnetic pressure, with matter accreting onto the central
PMS star in the disc mid-plane \citep{1996ARA&A..34..207H}. 

{\bf Metallic lines:} The spectrum of FU Ori contains absorption feature of the metallic lines of Na I at
2.206 $\mu$m. However, the metallic line of Ca at 2.256 $\mu$m is absent in our spectrum of FU Ori.
Previously, \citet{2018ApJ...861..145C} have reported weak presence of both these metallic lines in 
their spectrum of FU Ori. \citet{2009ApJ...693.1056L}, by their continuous NIR spectroscopic monitoring 
have shown that the metallic line of Na I likely originates in regions close to that of the CO bandheads and 
generally follows the profile of the latter, i.e. if the CO bandheads are in absorption, the Na I lines 
will be in absorption and vice versa. 

{\bf 1.083 $\mu$m He I:} In our spectrum of FU Ori, He I line at 1.083 $\mu$m  displays a prominent absorption feature. The absorption feature in He I is indicative of the outflowing winds generated due to accretion \citep{2003ApJ...599L..41E}. 

{\bf Ca II infra-red triplets:} The Ca II infra-red triplet (IRT) in the spectrum of FU Ori is found to be
in absorption. The absorption feature in the spectrum of FU Ori can be explained by the 
disk wind models of \citet{1990ApJ...349..168H}. These winds models describes about the various observed line profile shapes in terms of outflowing wind densities,
temperatures and velocity fields.

{\bf O I 7773 \AA~:} The O I line at 7773 \AA~ has a high ionisation state of $\sim$ 9 eV which prevents 
its formation in the photosphere of cool stars and it is believed to form in the broad component regions 
(BCRs) that correspond to the warm gas in the envelope and is therefore, believed to acts as a tracer for disc 
turbulence \citep{1992ApJS...82..247H}. This likely explains
the observed absorption feature at 7773 \AA.

{\bf K I:} We have also detected blue-shifted absorption feature in the metallic line of potassium,
K I at 7699 \AA. This line is indicator of the disc winds from FU Ori \citep{1998apsf.book.....H}.

{\bf H$\alpha$:} The H$\alpha$ profile exhibited by the spectrum of FU Ori is very complex. It displays primary peak along with a blue-shifted secondary peak which is greater than
the primary peak.
The observed line profiles can be explained in terms of the wind models of \citet{1990ApJ...349..168H}. 
The most likely wind models that can simultaneously explain the observed Ca II IRT and the H$\alpha$ are the 
Model 3(I) and Model 4(I) presented in Table 1 of \citet{1990ApJ...349..168H} where the regions from which 
these winds arise likely have temperatures ranging between
5000 K - 10000 K.

{\bf Na I D$_{1/2}$:} The spectrum of FU Ori also displays the absorption profile in the Na I D lines. With
the resolution of {\it TANSPEC} we can resolve the doublet lines of sodium. The absorption feature is typical
and is indicative of the outflowing winds. The observed
blue-shifted Na I D lines places a further qualitative constraint on the mass loss rate to be 10$^{-7}$ M$_{\odot}$/yr or greater \citep{1990ApJ...349..168H} 

\subsection{\bf XZ Tau} 

XZ Tau is a Classical T Tauri (CTT) star of spectral type M3 located in the L1551 dark cloud of Taurus star
forming region. It has been been classified as an EXor type variable by \citet{2009ApJ...693.1056L} based on 
its photometric and spectroscopic evolution. During our observation the brightness of XZ Tau was 
$\sim$ 13.2 mag in V band. This enabled us to obtain the spectrum of XZ Tau from {\it TANSPEC}. We can identify several spectroscopic features, some of which are in absorption while the
rest are in emission.  The main spectral features that we can identify are summarised as below: 

{\bf CO bandheads in K band:} We can identify very weak absorption CO overtone features of (2-0) and (3-1) 
transitions in our spectrum. Previously, \citet{2009ApJ...693.1056L} reported the bandheads to be
in absorption in their low resolution (R $\sim$ 250) spectra. This weakening of the CO overtone feature as 
observed in our spectra can be attributed to the change (increase) in the stellar temperature, resulting in increased irradiation of the inner disc, which decreases the strength of the CO bandheads (cf. Figure 7 of \citealt{1991ApJ...380..617C}). 
This variation in the CO bandhead features of months timescales was previously reported by \citet{2009ApJ...693.1056L}.

{\bf 2.2 $\mu$m Na I:} The atomic line Na I has an ionisation potential of 5.1 eV which is very low, hence
it is believed to have originated at the inner part of the circumstellar disc which are capable of 
shielding the ionising photons \citep{2009ApJ...693.1056L}. The Na I line in the spectrum of XZ Tau is 
found to be in absorption. This reinforces our understanding that the Na I and the CO bandheads have common origin \citep{1988ApJ...334..639M}.

{\bf 2.165 $\mu$m Br$\gamma$:} Br$\gamma$ line is an important tracer for magnetospheric accretion in the
NIR regime \citep{2018ApJ...861..145C}. Br$\gamma$ line in the spectrum of XZ Tau displays a P Cygni profile 
hence it can be assumed that the accretion process in XZ Tau is likely to be of magnetospheric origin along with outflowing winds from regions close to the accretion columns.

{\bf 1.28 $\mu$m Paschen$\beta$:} Hydrogen recombination line of Paschen$\beta$ is another diagnostic tracer of accretion as this line in the spectrum is believed to originate in the magnetospheric accretion columns  \citep{1997IAUS..182P.272F}. In the spectrum of XZ Tau we note that the Paschen$\beta$ line displays a P Cygni
profile which again can be attributed to the outflows originating from the regions close to the magnetospheric
accretion columns. 

{\bf 1.083 $\mu$m He I line:} The He I line at 1.083 $\mu$m is produced due to resonant scattering from the meta-stable Helium atoms and is an excellent diagnostic probe for the outflowing winds from YSOs 
\citep{2003ApJ...599L..41E}. The EUV to X-ray radiation formed due to the accretion results in 
the formation of the meta-stable triplet state of Helium atoms which then resonant scatter the 1.083 $\mu$m 
photons resulting in strong absorption signal in the spectrum. Our spectrum of XZ Tau exhibit absorption 
trough at 1.083 $\mu$m indicative of the outflows generated due to the magnetospheric accretion. Previously,
\citet{2017A&A...606A..48A} reported complicated profile of the He I line in their high resolution spectra. 
We have not observed any such complicated profile in our spectrum. The non-detection can either be attributed 
to the lower resolution of our spectrum being unable to resolve those features or to the change in the 
structure of the outflowing wind.

{\bf Ca II IRT line:} Emission feature in the infra-red triplet lines of Ca II in YSOs is thought to originate 
from the active chromospheres and the magnetospheric accretion funnels
\citep{1992ApJS...82..247H,1998AJ....116..455M}. In our spectra of XZ Tau, the Ca II IRT are
found to be in emission, indicative of the magnetospheric accretion process being the agent behind the accretion
of matter from the circumstellar disc to the central pre-main sequence star of XZ Tau. 

{\bf H$\alpha$:} The spectrum of XZ Tau exhibits an emission feature in H$\alpha$. This emission feature in
H$\alpha$ likely points towards a combination of accretion process, hot spots, magnetic field topology etc., 
in XZ Tau \citep{2014A&A...561A...2A}.

\subsection{\bf V2492 Cyg}

V2492 Cyg also known as PTF 10nvg is a Class {\sc i} pre-main sequence star that underwent its first recorded
outburst in 2010 \citep{2011AJ....141..196A}. It has displayed a highly variable photometric behaviour since then, with it transitioning to outburst state in 2016-2017 \citep{2017ATel10183....1M}. It has continued to 
show its highly variable behaviour after this latest episode of outburst. 
The physical origins of the variability observed in V2492 Cyg has been attributed to be driven by both UXor-type extinction\footnote{UXor-type variability arises due to occultation of the central YSO by the matter present in the circumstellar disc} 
and EXor-type  accretion \citep{2013AJ....145...59H,2018A&A...611A..54G,2021BlgAJ..35...54I}.  
A paper describing about the detailed evolution of the various spectral features from previous outbursts till present times is currently under preparartion.
The spectral lines that we have obtained from our {\it TANSPEC} spectrum of V2492 Cyg are 
the CO bandhead features in K band, 2.2 $\mu$m Na I, 2.165 $\mu$m Br$\gamma$, 1.28 $\mu$m Paschen$\beta$,
1.083 $\mu$m He I, Ca II IRT and H$\alpha$.

\subsection{\bf V2494 Cyg}

V2494 Cyg is located in the L1003 dark cloud of the Cygnus OB7 star forming complex at a distance of 800 pc \citep{2013MNRAS.432.2685M}. V2494 Cyg was spectroscopically classified as an FUor source owing to its similarity with the spectrum of FU Ori by \citet{1997AJ....114.2700R}. The source was faint during our 
observation with a magnitude of $\sim$ 17.5 mag in $zr$ band of Zwicky Transient Facility 
(ZTF) and therefore we could only obtain the NIR spectrum of V2494 Cyg. 

The prominent lines visible in our spectra are the CO bandhead features in K and Paschen$\beta$. A detailed analysis of the spectral line features and their evolution will
be described in a separate upcoming paper.



\subsection{\bf V1180 Cas}

V1180 Cas is variable star located in the dark cloud Lynds 1340 in the star forming region in Cassiopei, 
located at a distance of 600 pc from the Sun. It was identified by \citet{1994A&A...292..249K} as a variable young star based on its strong H$\alpha$ emission. V1180 Cas, similar to V2492 Cyg exhibits highly variable 
photometric behaviour that can be attributed to a combination of extinction and accretion rate changes \citet{https://doi.org/10.48550/arxiv.2210.09660}. The main spectral features of V1180 Cas as obtained from {\it TANSPEC} are described below:

{\bf CO bandheads in K:} The CO bandheads in the spectrum of V1180 Cas are in emission implying the strong 
irradiation from the central pre-main sequence star to the inner part of the circumstellar disc \citep{1991ApJ...380..617C}. 

{\bf 2.165 $\mu$m Br$\gamma$:} In the NIR regime of the spectrum of T Tauri stars, Br$\gamma$ is an important 
signature of the magnetospheric accretion \citep{2018ApJ...861..145C}. The spectrum of V1180 Cas also exhibits a strong Br$\gamma$ emission line.

{\bf Hydrogen recombination lines of Brackett series in H band:} The spectrum of V1180 Cas several Brackett series hydrogen recombination lines with all of them being found to be in emission. We can identify the Brackett (12-4), Brackett (11-4) and Brackett (10-4) recombination lines. 

{\bf 1.28 $\mu$m Paschen$\beta$:} The Paschen$\beta$ line in the spectrum of V1180 Cas is in emission. The Paschen$\beta$ line profile observed in V1180 Cas is similar to that observed in the case of V2492 Cyg implying similar physical processes governing its evolution. 

{\bf 1.083 $\mu$m He I:} The He I line at 1.083 $\mu$m exhibits a P Cygni profile in the spectrum of V1180 Cas. P 
Cygni profile as observed in the spectrum therefore signifies the heavy mass loss via outflowing winds from 
the innermost regions of the disc from where the accretion onto the PMS star is occurring \citep{2003ApJ...599L..41E}.   

{\bf Ca II IRT lines:} The Ca II IRT lines are found to be in emission therefore indicating magnetospheric
accretion of matter from the circumstellar disc. 

{\bf H$\alpha$:} The H$\alpha$ line in the spectrum of V1180 Cas is in emission and is considerably stronger 
in strength compared to that observed in V2492 Cyg and XZ Tau. 

\subsection{\bf V899 Mon}

V899 Mon is located in the Monocerous R2 region of our galaxy at a distance of about 905 pc \citep{2011A&A...535A..16L}. It was reported as an episodic accretion candidate by the Catalina Real Time
Survey (CRTS) when it underwent an outburst in 2010 and subsequent spectrum of the source displayed 
strong H$\alpha$ and Ca II IR triplet lines which confirmed it to be an YSO \citep{2009CBET.2033....1W}.
V899 Mon is peculiar kind of YSO exhibiting episodic accretion behaviour. During its 2014 outburst, 
V899 Mon displayed properties intermediate to that of FUors and EXors \citep{2015ApJ...815....4N}. 
Spectroscopically, it displayed P Cygni profile in H$\alpha$ and  Ca II IRT, while the prominent NIR 
lines (1.083 $\mu$m He I and CO bandheads in K band) were found to be in absorption. After the 2014 outburst, 
V899 Mon has been steadily transitioning towards its quiescent state. Recent spectroscopic observations 
made by \citet{2021ApJ...923..171P} reveals transition of the P Cygni profile of H$\alpha$ and Ca II IRT into 
emission feature. The NIR CO bandhead features also transitioned into emission.   
 We will be discussing about the detailed evolution of the various spectral features and the likely physical processes governing their evolution in a separate paper that is currently under preparation.
The prominent spectral features of V899 Mon that we have observed with {\it TANSPEC} are
CO bandheads in K, 2.165 $\mu$m Br$\gamma$, several Hydrogen recombination lines in H band like Br10 at 1.736 $\mu$m, Br11 at 1.681 $\mu$m, Br12 at 1.641 $\mu$m, B13 at 1.611 $\mu$m, Br15 at 1.571 $\mu$m, Br16 at 1.556$\mu$m and Br17 at 1.544 $\mu$m. 
Several Paschen series lines were also detected in the spectrum like the Paschen$\beta$,
Paschen$\gamma$ at 1.094 $\mu$m and Paschen$\delta$ at 1.005 $\mu$m.
We have also detected the  1.083 $\mu$m He I, the IR triplet of Ca II and H$\alpha$ in the spectrum.

\subsection{\bf V960 Mon}

V960 Mon is also known as 2MASS J06593158-0405277 is located in the Monocerous star forming region of the 
galaxy at a distance of 450 pc. V960 Mon underwent its first recorded outburst in 2014 and has since been 
steadily fading from its peak brightness since then, typical of sources displaying FUOrionis behaviour \citep{2016NewA...43...87J}. The spectral characteristics exhibited by V960 Mon was also similar to that
of FU Ori, the first member of the FUOrionis subclass of episodically accreting young stars \citep{2014ATel.6797....1H,2015ATel.6862....1R}. V960 Mon was, spectroscopically monitored after its 
2014 outburst by \citet{2018AJ....155..101T}. Their monitoring revealed gradual decline in the luminosity of 
the accretion disc post outburst due to decrease in accretion rate. Recently,  \citet{2020ApJ...904...53T} and \citet{2020ApJ...900...36P} presented further spectroscopic monitoring results of V960 Mon. Their results indicate a gradual weakening of the wind features post outburst along with the presence of
double peaked line profiles indicative of Keplerian rotation. Spectroscopic monitoring further revealed that the spectrum of V960 Mon has become central star dominant.

The main spectral features that we have observed  with our {\it TANSPEC} spectrum are the CO bandheads in K, 2.12 $\mu$m H$_2$ line, 1.28 $\mu$m Paschen$\beta$ line, 1.083 $\mu$m He I and the H$\alpha$. A detailed analysis of the observed line profiles of V960 Mon and the
possible underlying physical processes will be described in an upcoming paper.

\subsection{\bf V1118 Ori}

V1118 Ori has been classified belonging to the EXor subgroup of eruptive young variables. It underwent its
first recorded outburst in 2006 after which it remained in quiescence for almost a decade, following which
it again underwent an outburst during 2015-2016 period \citep{2017ApJ...839..112G}. V1118 Ori again
underwent an outburst during 2019 which lasted almost for a year and by February 2020 it had transitioned
back to its quiescent phase \citep{2020Ap....tmp...75G}. We have obtained our spectrum of V1118 Ori on 
2020 October 29. In the following, we will discuss about the spectral line profiles that we have obtained.

{\bf CO bands in K:} In our spectrum of V1118 Ori, the CO bandheads are found to display a weak absorption 
profile. Previously, \citet{2015ApJ...802...24L,2016ApJ...819L...5G, 2017ApJ...839..112G} have reported 
the CO bandheads to be in emission. The weak bandheads possibly hints towards the change in the inner 
structure of the circumstellar disc as the CO bandheads in emission is thought arise from the warm gas in
the innermost part of the disc powered by active accretion
\citep{1996ApJ...462..919N,2009ApJ...693.1056L,2011ApJ...736...72K}. In the case of XZ Tau 
the presence of 2.2 $\mu$m Na I in absorption helped to definitively conclude about the weak absorption in the CO bandheads. However, the spectrum of V1118 Ori lacks the Na I line hence we refrain from inferring
anything from the observed bandhead features, rather more time-series spectroscopy of this source is required
to understand the evolution of CO bandheads.

{\bf 2.165 $\mu$m Br$\gamma$:} The Br$\gamma$ line our spectrum of V1118 Ori is found to display a P Cygni
profile. Presence of P Cygni profile in Br$\gamma$ points towards heavy outflow from from the regions 
close to accretion as Br$\gamma$ is an important tracer of accretion in the NIR \citep{2018ApJ...861..145C}.

{\bf H$_2$ lines in K:} We have detected the H$_2$ (1-0) transition feature to be in emission in our
spectrum of V1118 Ori. H$_2$ lines are absent in the spectrum of FUors with V960 Mon to be an exception
\citet{2020ApJ...900...36P}. However, H$_2$ lines are seen in several EXors, like ESO$-$H$\alpha$ 99 \citet{2019AJ....158..241H}, V346 Nor \citet{2020ApJ...889..148K}, Gaia 19bey \citet{2020AJ....160..164H}
 etc. Emission feature in H$_2$ (1-0) is indicative of the outflows in young stars
\citep{2003A&A...397..693D,2010A&A...511A..24D,2007prpl.conf..215B}. 

{\bf Hydrogen recombination lines of Paschen series:} The hydrogen recombination lines in emission are an 
important diagnostic
tracer of the magnetospheric accretion \citep{1997IAUS..182P.272F}. In our spectrum of V1118 Ori, 
Paschen$\beta$ and Paschen$\gamma$ line displays a prominent emission profile thereby indicating 
magnetospheric accretion to be main accretion process in V1118 Ori.

{\bf 1.083 $\mu$m He I:} The 1.083 $\mu$m He I line in our spectrum of V1118 Ori is found to be in emission. The 
emission feature we have observed, has been seen previously in the spectrum of V1118 Ori, as reported by 
\citet{2015ApJ...802...24L} during the decades long quiescence of V1118 Ori and also during its 2015 outburst
\citep{2016ApJ...819L...5G}. He I has a very high ionisation potential and its presence in emission can be likely
attributed to composite origin arising from accretion funnel flow and/or from an accretion shock \citep{2003ApJ...599L..41E}

{\bf Ca II IRT:} The emission features in Ca II IRT lines are believed to form in the magnetospeheric accretion 
funnels in YSOs \citep{1998AJ....116..455M}. In our spectrum of V1118 Ori, we have found the Ca II IRT lines to 
be in emission thereby likely indicating of magnetospheric accretion process. 

{\bf H$\alpha$:} The H$\alpha$ in emission in YSOs likely arises because of several processes like chromospheric
activity, hot spots on the stellar surface, stellar rotation, magnetic field topology, etc. apart from the 
accretion rate \citep{2014A&A...561A...2A}. In our spectrum of V1118 Ori, H$\alpha$ line is found to be in 
emission. As V1118 Ori is currently in quiescent phase, therefore, the likely significant contribution to the 
emission line in H$\alpha$ is from the other processes as mentioned above apart from accretion rate.

\subsection{\bf V733 Cep}

V733 Cep, also known as the Persson's star is located in the L1216 = Cep F cloud in the Cepheus region at a 
distance of 800 pc \citep{2007AJ....133.1000R}. Based upon the Spectral Energy Distribution (SED) analysis and
spectroscopic similarity with that of FU Ori, \citet{2007AJ....133.1000R} classified V733 Cep as an FUorions 
candidate, with it having undergone an outburst sometime between 1953 and 1984. 
The spectral features that we have observed in our {\it TANSPEC} spectrum of V733 Cep are
CO bandheads in K, H$_2$ 1-0 S(10) at 1.6665 $\mu$m, 1.28 $\mu$m Paschen$\beta$ 
and the 1.083 $\mu$m He I. We are currently studying the evolution of the
various spectral features of V733 Cep in detail and a separate paper is currently under
preparation.





\vspace{2em}
\section{Summary \& Conclusion} \label{pt4}

We have presented here a single epoch spectra of 9 young eruptive variables that are either classified as FUor or EXor. Our observations with {\it TANSPEC}
clearly emphasises the unique wide spectral coverage of {\it TANSPEC} starting at optical wavelengths of 
0.55 $\mu$m to all the way till the NIR wavelengths at 2.5 $\mu$m. The resolution and 
especially the wide spectral range makes {\it TANSPEC} an ideal instrument to probe the innermost regions
of the circumstellar disc of YSOs from where the accretion and the outflow process onto the young star
occurs. In this paper, we have analysed the line profile shapes of the observed set of episodically accreting
young stars. We have tried to explain the observed line profile shapes based on the existing models that explains the physical processes
that occur in these group of sources. The episodically accreting stars are broadly classified into FUors
and EXors with our current theoretical understanding is that the accretion mechanism onto the central
young star is possibly different in the subgroups \citep{1996ARA&A..34..207H}. Our observation of the 
set of FUors and EXors points to the similar understanding, however, more spectroscopic observations with a
larger set and time series spectra are needed to validate the current theoretical understanding.

The main conclusions that can be drawn from this study are summarised below: 

\begin{enumerate}
  
  \item The classical bi-modal spectroscopic classification of the episodically accreting young stars based
  on the absorption/emission profile of the CO bandheads probably needs a revision. For better 
  categorisation, we need to look at a set of common line profile shapes and compare them with the
  current theoretical models.
  
  \item The changing of the spectral line profiles and emergence of new spectroscopic features, as observed
  in the cases of XZ Tau and V1118 Ori brings forth the necessity of the long term spectroscopic
  monitoring for better understanding of the physical processes that govern the disc evolution and the 
  accretion dynamics. Time evolution spectroscopic studies will also help to establish link between the
  accretion and outflow processes.
  
  \item We further identify V960 Mon, which is classified as FUor, to be somewhat peculiar in the sense that it displays spectroscopic features like the H$_2$ recombination lines in its spectrum that are generally associated with the spectrum of EXors.
  
  \item We plan to further monitor the source V1118 Ori spectroscopically. The CO bandhead features were 
  prominently detectable in emission in its spectrum. However, in our spectrum, the features were barely 
  detectable. Therefore, we urge for more observations to better understand its current evolution and also
  to corroborate our observation.
  
  \end{enumerate}



\section*{Acknowledgements}
\vspace{1em}

We thank the anonymous reviewer for valuable comments which greatly improved the scientific content of the 
paper. TIFR$-$ARIES Near Infrared Spectrometer (TANSPEC) was built in collaboration 
with TIFR, ARIES and MKIR, Hawaii for the DOT. We thank the staff at the 3.6m DOT, Devasthal (ARIES), for their co$-$operation during observations.
It is a pleasure to thank the members of 3.6m DOT team and IR astronomy group at TIFR for their support
during {\it TANSPEC} observations. SS  acknowledge  the support of the Department of Science  and
Technology,  Government of India, under project No. DST/INT/Thai/P-15/2019. DKO acknowledges the support of
the Department of Atomic Energy, Government of India, under Project Identification No. RTI 4002. JPN and DKO 
acknowledge the support of the Department of Atomic Energy, Government of India, under project Identification
No. RTI 4002.

\bibliography{tanspec_spectra_atlas}{}




\end{document}